\pdfoutput=1
\documentclass[11pt]{article}
\usepackage[]{acl}
\usepackage{times}
\usepackage{latexsym}
\usepackage[T1]{fontenc}
\usepackage[utf8]{inputenc}
\usepackage{microtype}
\usepackage{booktabs}
\usepackage{multirow}
\usepackage{graphicx}

\usepackage{amsmath,amsfonts,bm}

\def\eqref#1{equation~\ref{#1}}

\def\1{\bm{1}}

\def\rd{{\textnormal{d}}}

\def\rx{{\textnormal{x}}}

\def\vp{{\bm{p}}}

\def\vu{{\bm{u}}}
\def\vv{{\bm{v}}}

\DeclareMathAlphabet{\mathsfit}{\encodingdefault}{\sfdefault}{m}{sl}
\SetMathAlphabet{\mathsfit}{bold}{\encodingdefault}{\sfdefault}{bx}{n}

\def\gD{{\mathcal{D}}}

\def\gM{{\mathcal{M}}}

\def\gO{{\mathcal{O}}}

\def\sD{{\mathbb{D}}}

\def\sN{{\mathbb{N}}}

\newcommand{\R}{\mathbb{R}}

\newcommand{\softmax}{\mathrm{softmax}}

\usepackage{mathtools}
\usepackage{multirow}
\usepackage{multicol}
\usepackage{booktabs}
\usepackage{subfigure}

\DeclareMathOperator*{\enc}{Enc}

\DeclareMathOperator*{\transformerenc}{Transfm-Enc}

\DeclareMathOperator*{\transformercls}{Transformer-cls}

\DeclareMathOperator*{\ffn}{FFN}

\DeclareMathOperator*{\kldiv}{KL-Div}

\title{Fine-Grained Distillation for Long Document Retrieval}

\author{Yucheng Zhou$^1$\thanks{~~Work done during the internship at Microsoft.}, Tao Shen$^2$,  Xiubo Geng$^2$, Chongyang Tao$^2$, \\ { \bf Guodong Long$^1$, Can Xu$^2$, Daxin Jiang$^2$\thanks{~~Corresponding author.} }
\\
  $^1$AAII, School of CS, FEIT, University of Technology Sydney \\
  $^2$Microsoft Corporation\\
{\tt yucheng.zhou-1@student.uts.edu.au,guodong.long@uts.edu.au} \\ 
{\tt \{shentao,xigeng,chotao,caxu,djiang\}@microsoft.com}
}

\begin{document}
\maketitle
\begin{abstract}
Long document retrieval aims to fetch query-relevant documents from a large-scale collection, where knowledge distillation has become de facto to improve a retriever by mimicking a heterogeneous yet powerful cross-encoder. However, in contrast to passages or sentences, retrieval on long documents suffers from the \textit{scope hypothesis} that a long document may cover multiple topics. This maximizes their structure heterogeneity and poses a granular-mismatch issue, leading to an inferior distillation efficacy. In this work, we propose a new learning framework, fine-grained distillation (FGD), for long-document retrievers. While preserving the conventional dense retrieval paradigm, it first produces global-consistent representations crossing different fine granularity and then applies multi-granular aligned distillation merely during training. In experiments, we evaluate our framework on two long-document retrieval benchmarks, which show state-of-the-art performance.
\end{abstract}

\section{Introduction}
Large-scale retrieval, as a fundamental task in information retrieval (IR), has attracted increased interest from industry and academia in the last decades, as it plays an indispensable role in a wide range of real-world applications, such as web engines \cite{Fan22Pre}, question answering \cite{Karpukhin2020DPR} and dialogue systems \cite{Yu21Few}.
Given a text query, it aims to fetch top-relevant documents\footnote{Each entry of the collection can be any text granularity (e.g., sentence, passage, document) but we take `document' to denote `entry of collection' in this paper for clear writing.} from a huge collection \cite{Cai2021IRSurvey}.
As the collection usually scales up to millions or billions, a retrieval method must satisfy the efficiency or latency requirement of online deployment to calculate the relevance score between a query and every document. 

\begin{figure}[t]
    \centering
    \includegraphics[width=\linewidth]{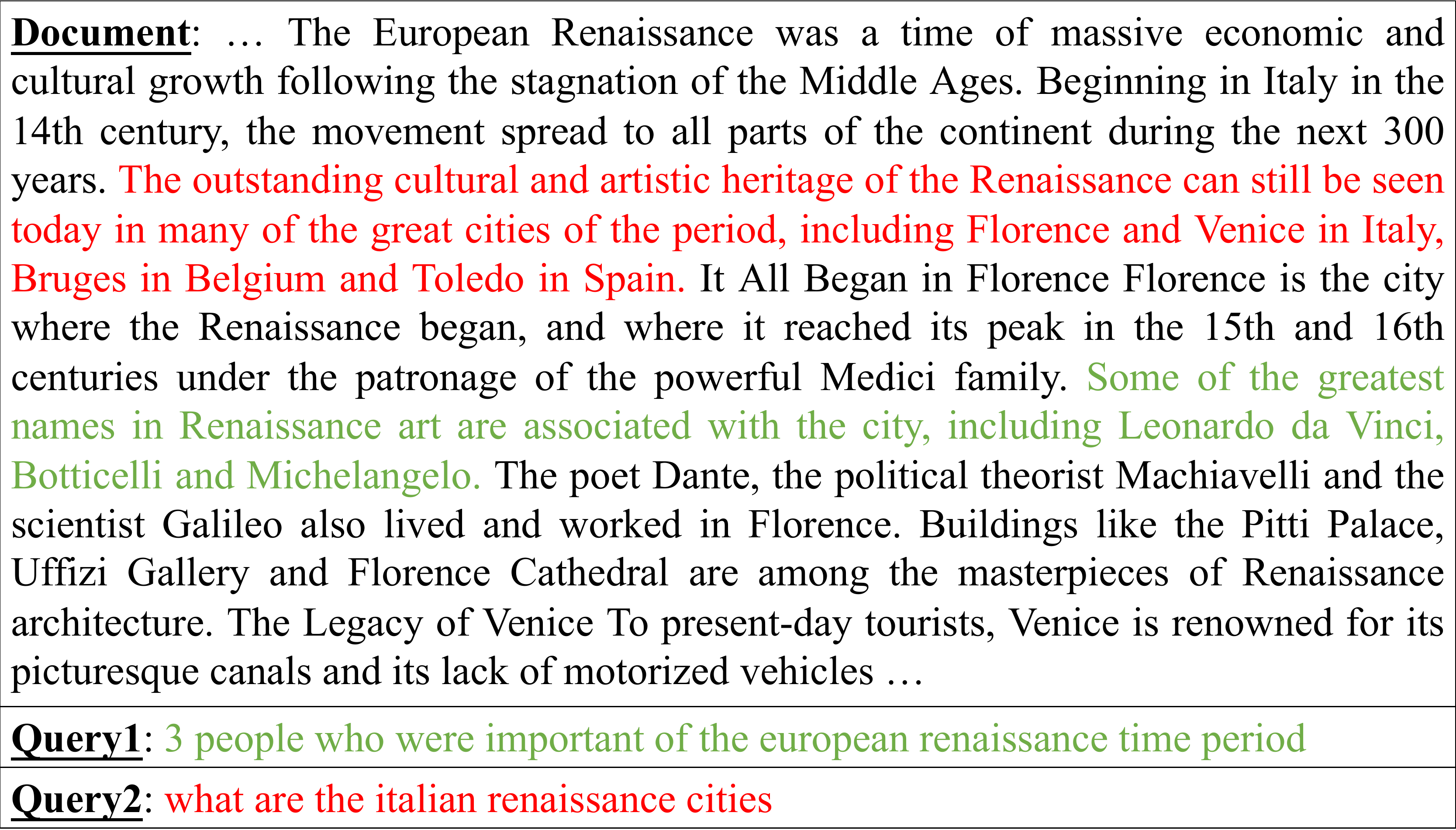}
    \caption{\small A case for \textit{scope hypothesis} in long document.}
    \label{fig:intro}
\end{figure}

Recently, pre-trained language models (PLMs), e.g., BERT \citep{Devlin2019BERT}, RoBERTa \citep{Liu2019RoBERTa}, DeBERTa \citep{He2021DeBERTa}, have dominated the field of IR in deep representation learning literature, as they are readily adapted to capture token-wise correlations and produce generic representations by fine-tuning. 
In the common practice of PLMs, a pair of text pieces (i.e., a query and every document in our task) should be concatenated to pass into the models \citep{Devlin2019BERT} for fine-grained relevance measurement -- known as \textit{cross-encoder} that performs very competitively -- however cannot meet the efficiency requirement due to combinatorial explosion in terms of online PLM inference \citep{Zhang2021AR2,Ren2021RocketQAv2}. 
In contrast, a \textit{bi-encoder} (a.k.a. dual-encoder or two-tower) leverages the PLMs to embed queries and documents individually into a single vector in the same dense semantic space, and then query-document relevance can be derived by a lightweight metric (e.g., dot-product)  \citep{Reimers2019SentenceBERT}. 
The bi-encoder enables offline document embeddings and satisfies the online efficiency requirement, so it has become the de facto model choice for PLM-based large-scale retrievers. However, the bi-encoder is vulnerable to information bottleneck by the single dense vector and thus lags behind the cross-encoder considerably  \citep{Wang2022simlm,Gao2022coCondenser,Liu2022RetroMAE}.

To narrow the performance gap against cross-encoder, a recently advanced technique to train bi-encoder is distilling list-wise relevance score distributions from cross-encoder during contrastive learning \citep{Zhang2021AR2,Ren2021RocketQAv2}. 
This technique merely affects the training process of a bi-encoder and has been proven to improve the generalization ability of bi-encoder \citep{menon2021indefense}, leading to better retrieval quality without any sacrifice of inference efficiency.

Nonetheless, such a distillation technique to improve bi-encoder has proven effective merely in the scenarios where the targeted text pieces are usually short semantic units (e.g., sentences \citep{Liu2022transenc} and passages \citep{Ren2021RocketQAv2}) with an almost single topic. 
In contrast, long document retrieval usually targets super-long documents with up to thousands of words (cf. 65 words per passage \citep{Nguyen2016MSMARCO}). 
Considering the \textit{scope hypothesis} \citep{Robertson09Probabilistic} that a long document may cover multiple topics (see a case in Figure~\ref{fig:intro}), distilling knowledge from a cross-encoder to bi-encoder is prone to become less effective. 
This is likely because modeling long documents maximizes their heterogeneity in terms of visibility
-- cross-encoder explicitly models the query-dependent salience part (e.g., a sentence) whereas bi-encoder directly models the whole into a query-agnostic dense bottleneck
-- thus such a brute-force distillation suffering from the granularity mismatching. 
In our pilot experiments, the brute-force distillation can only bring $0.1\%$ gain on long document retrieval after extensive tuning, in contrast to $>1\%$ gain frequently observed in passage retrieval \citep{Ren2021RocketQAv2}.

Thereby, we aim to improve the knowledge distillation from a cross-encoder to a long-document retriever by circumventing the granularity mismatching problem. 
Instead of knowledge distillation at the long-document level, we propose a brand-new bi-encoder learning framework, dubbed fine-grained distillation (FGD), for large-scale retrieval over long documents. 
Basically, it operates on multi-vector distillation crossing fine granularity merely in the training phase while keeping single-vector retrieval during inference. 
To derive fine-grained representations without cross-granular conflict, we first propose a global-consistent granularity embedding method, which enables dynamic contextualization visibility (e.g., passage, sentence) over a long document. 
Then, we present a local-coordinating score distilling strategy, which replaces global (i.e., document-level) distillation, for long-document retriever training. 
In addition, to empower our distillation strategy, we further propose a hierarchical negative mining technique to produce hard negatives throughout granularity. 

In the experiments, we conduct an extensive evaluation of our proposed framework on two document retrieval benchmark datasets, i.e., MS-Marco document retrieval \cite{Nguyen2016MSMARCO} and TREC 2019 Deep Learning track \cite{Craswell2020TREC19}. 
The experimental results show that our method achieves state-of-the-art performance compared with other strong competitors. 
In addition, we verify the generality of our framework by evaluating it on different long document retrievers paired with different cross-encoder teachers.

\section{Methodology}

\begin{figure*}[t]
    \centering
    \includegraphics[width=0.8\linewidth]{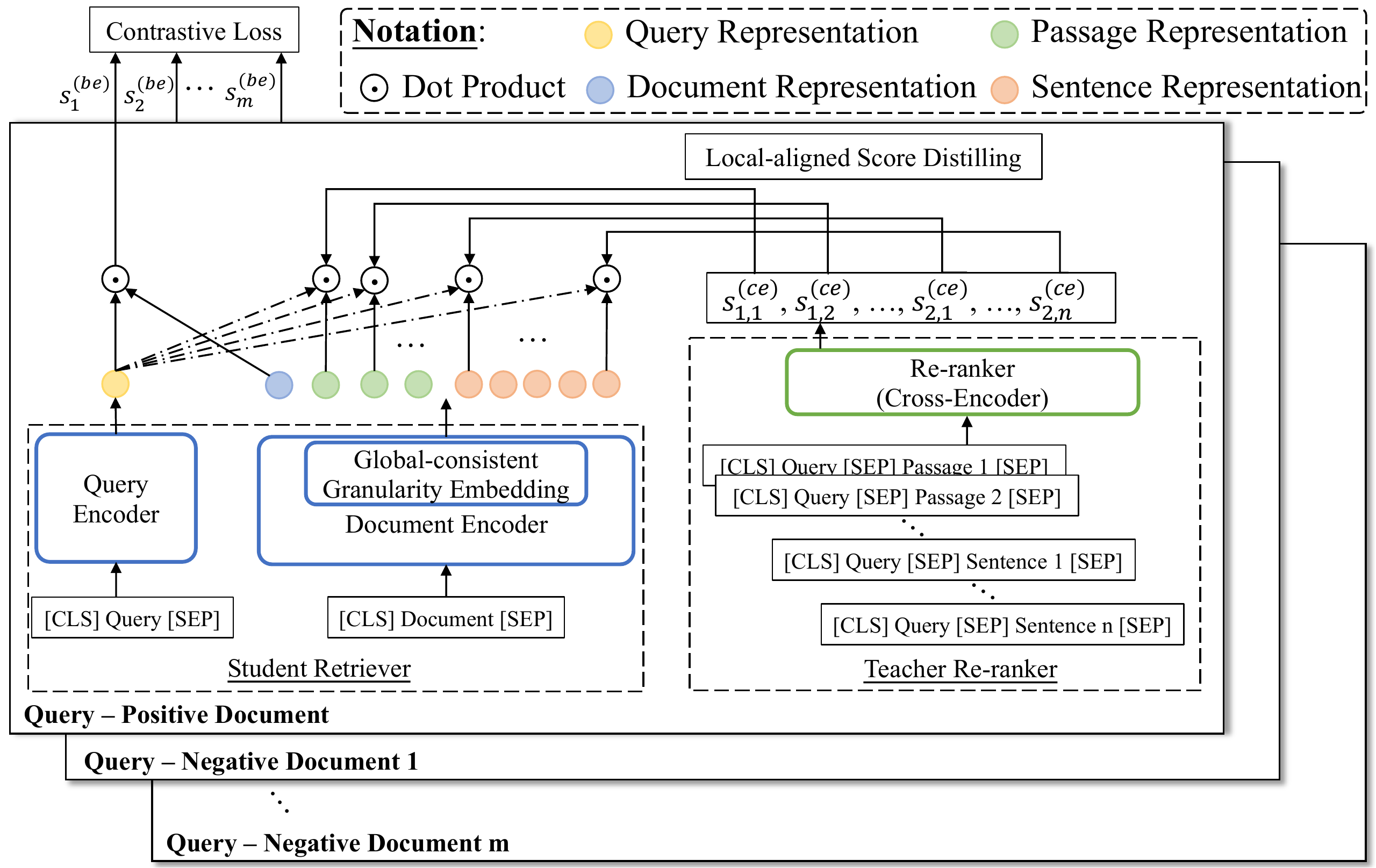}
    \caption{\small An overview of our fine-grained distillation (FGD) for long-document retrieval.}
    \label{fig:model}
\end{figure*}

\paragraph{Task Definition.} Considering a large-scale collection with numerous long documents (i.e., $\sD = \{d_i\}_{i=1}^{|\sD|}$ where each $d_i$ denotes a document), large-scale retrieval is to fetch top-relevance documents (i.e., $\bar{\sD}^q$) by a retriever (e.g., $\gM$) for a text query $q$. 
This requires $\gM$ to calculate every relevance score $s^q_i$ between the $q$ and $\forall d_i\in\sD$, where $i\in [1,|\sD|]$. 
In the remaining, we will omit the superscript `$q$' for clean demonstration if no confusion is caused.

\subsection{Bi-encoder Learning with Distillation} \label{sec:be_learning}

To meet the efficiency requirement of large-scale retrieval, a de facto scheme \citep{Gao2021Condenser,Gao2022coCondenser,Wang2022simlm} is to leverage a bi-encoder for the relevance score. It encodes each query and document individually into dense semantic space and derives the score usually by a lightweight metric (e.g., dot-product, cosine similarity), This can be formally written as
\begin{align}
    s&^{(\text{be})} \coloneqq \gM^{(\text{be})}(q, d|\theta^{(\text{be})}) = <\vu,\vv> \coloneqq \label{equ:bi_enc} \\
    \notag &<\enc(q|\theta^{(\text{q})}), \enc(d|\theta^{(\text{d})})>,\exists d\in\gD,
\end{align}
where $<\cdot,\cdot>$ denotes a non-parametric dot-product, $\enc(\cdot|\theta^{(\text{*})})$ denotes a $\theta^{(\text{*})}$-parameterized encoder that embeds a piece of text into a dense vector, and $\theta^{(\text{be})} = \theta^{(\text{q})} \cup \theta^{(\text{d})}$ parameterize the bi-encoder where the query and document encoders can be tied in terms of parameters.

Then, the training of retrieval-related models (e.g., bi-encoder learning $\theta^{(\text{be})}$ here) is usually formulated as a contrastive learning problem. 
That is, only a positive document $d_+$ is given as a golden label for the query $q$, while a set of negative documents $d_-\in\sN$ also should be mined in light of ($q$, $d_+$) for contrastive learning \citep{Gao2022coCondenser}. 
Basically, a BM25 system or a trained retriever is usually employed to mine the negatives. 
Providing $d_+$ and $\sN$, we can derive a score distribution over them, i.e., 
\begin{align}
    \vp^{(\text{be})} &\coloneqq P(\rd|q, \{d_+\}\cup\sN;\theta^{\text{(be)}}) = \label{equ:bi_score_dist} \\
    \notag &\dfrac{\exp(\gM^{(\text{be})}(q, d|\theta^{(\text{be})})/\tau)}{\sum\nolimits_{d'\in\{d_+\}\cup\sN} \exp(\gM^{(\text{be})}(q, d'|\theta^{(\text{be})})/\tau)},
\end{align}
where $\forall d\in\{d_+\}\cup\sN$ and $\tau$ denotes the temperature set to $1$. Next, the training loss of contrastive bi-encoder learning can be simply written as
\begin{align}
    \notag L^{\text{(cl)}} &= - \sum\nolimits_q \log P(\rd=d_+|q, \{d_+\}\cup\sN;\theta^{\text{(be)}}) \\
    &= - \sum \log \vp^{(\text{be})}_{[\rd=d_+]}. \label{equ:contrastive_learning}
\end{align}

To improve the bi-encoder's generalization ability and boost its retrieval qualities, a common practice is to distill score distributions from a cross-encoder to the bi-encoder retriever.
In general, a cross-encoder is frequently defined as a Transformer-based classifier that a Transformer encoder followed by a one-way-out multi-layer perceptron (MLP). Hence, a cross-encoder can be formulated as
\begin{align}
    &s^{(\text{ce})} \coloneqq \gM^{(\text{ce})}(q, d|\theta^{(\text{ce})}) = \label{equ:cross_enc} \\
    \notag ~&\transformercls(\texttt{[CLS]} q \texttt{[SEP]} d \texttt{[SEP]}|~\theta^{(\text{ce})}), 
\end{align}
where $s^{(\text{ce})}\in\R$ and $\theta^{(\text{ce})}$ parameterizes this cross-encoder. 
Here, $q$ and $d$ concatenated with special tokens are passed into the self-attention encoder to enable token-level interaction, capture fine-grained nuance, and produce precise relevance score. 
Note that, $\theta^{(\text{ce})}$ can be either well-trained in advance \citep{Gao2022coCondenser,Zhou2022r2anker} or updated along with the bi-encoder \citep{Ren2021RocketQAv2,Zhang2021AR2}, while we opt for the former but without loss of generality. 
Next, we can also obtain $\vp^{(\text{ce})} \coloneqq P(\rd|q, \{d_+\}\cup\sN;\theta^{\text{(ce)}})$ as in Eq.(\ref{equ:bi_score_dist}).
Lastly, the loss function of such distillation is 
\begin{align}
    L^{\text{(kd)}} = \kldiv(\vp^{(\text{be})}\Vert\vp^{(\text{ce})}).
\end{align}

So, the final training loss for the bi-encoder learning with distillation is written as $\lambda L^{\text{(cl)}} + L^{\text{(kd)}}$.

\subsection{Global-consistent Granularity Embedding}

Although the bi-encoder learning with distillation has been proven very effective in passage retrieval \citep{Wang2022simlm} or sentence matching \citep{Reimers2019SentenceBERT}, its efficacy will be diminished when directly applied to long-document retrieval due to granularity mismatch. 
This is because the cross-encoder defined in Eq.(\ref{equ:cross_enc}) is able to focus only on the $q$-relevant topic of $d$ via its fine-grained self-attention mechanism, regardless of other topics in the scope hypothesis. 
By comparison, the bi-encoder defined in Eq.(\ref{equ:bi_enc}) is constrained by its representation bottleneck (i.e., fixed-length low-dimensional vector by $\enc(\cdot)$), so it can only produce $q$-agnostic $d$ representations as a whole. 

To break the bottleneck during distillation, we propose to perform knowledge distillation over fine-grained text pieces instead of the whole document. 

However, an open question remains about how to derive consistent embeddings across granularity. 
In particular, to produce consistent embeddings, previous methods directly apply mean-pooling over contextual embeddings for different granularity, which however becomes inferior when the document length goes extremely long and has proven less effective in our pilot experiments. This is the reason why most previous document retrieval works rely on \texttt{[CLS]} embedding paradigm \citep{Ma2022COSTA,Xiong2021ANCE,Zhan2021STAR-ADORE,Lu2021SeedEncoder}. 

Thereby, to better align with the prevalent \texttt{[CLS]} embedding paradigm, we present a global-consistent granularity embedding method. 
Specifically, `\texttt{[CLS]} embedding' denotes using the contextual embedding of \texttt{[CLS]} to represent the whole sequence, which is equivalent to applying a self-attention pooling \citep{Lin2017structured-self-att,shen2018disan} to the penultimate layer, i.e.,
\begin{align}
    \notag \vv^d &= \transformerenc(\texttt{[CLS]}d\texttt{[SEP]}|\theta^{\text{(d)}})_{\texttt{[CLS]}} \\
    & = \ffn(\sum\nolimits_{i\in [1,|d|]} \sigma(\alpha_{\texttt{[CLS]}\leftarrow d_i}) h'_i )
\end{align}
where $i$ denotes the token index in $d$, $h'_i$ denotes a hidden state for token $d_i$ from the previous layer, $\sigma$ denotes a non-linear function and usually $\softmax$, $\alpha_{\texttt{[CLS]}\leftarrow d_i}$ denotes an attention probability from \texttt{[CLS]} to $d_i$, and $\ffn$ denotes post-processes including MLP and residual connection defined in the Transformer.
The attention scores are calculated between global embedding $h'_{\texttt{[CLS]}}$ and each token embedding $h'_i$ by the attention module in the last layer of the Transformer \citep{Vaswani2017Transformers}. 
Then, following such global-aware attention pooling, we can leverage the off-the-shelf attention scores to produce global-consistent embeddings across granularity. 
Formally, given an arbitrary text span $x \in d$ with the token indices $[b^x,e^x]$, its global-consistent embedding can be written as
\begin{align}
    \notag \vv^x & = \enc(x|d;\theta^{\text{(d)}}) \\
    & \coloneqq \ffn(\sum\nolimits_{i\in [b^x,e^x]} \sigma(\alpha_{\texttt{[CLS]}\leftarrow d_i}) h'_i). \label{equ:multi_granular_repre}
\end{align}
Consequently, we can readily derive representation for various granularity, e.g., passages and sentences, via $\enc(x|d;\theta^{\text{(d)}})$. 

\paragraph{Remark on Embedding Propagation. }
In addition to the aforementioned mean-pooling methods \citep{Reimers2019SentenceBERT}, a recent trend to get multi-granular representation is employing graph neural network (GNN) \citep{wu2021gnn4nlp} for deep embedding propagation \citep{Zheng2020DocGNN}. 
Both of them focus on fine-grained representations rather than document-level ones and target the final applications of the representations, e.g., open-domain and context-based question answering. 
Standing with a distinct motivation, we still focus on the single document-level bottleneck but leverage fine-grained representations as the intermediate for knowledge distillation. 
This necessitates the paradigm of original global \texttt{[CLS]} representation, which requires consistency between document-level and fine-grained representations without complicated embedding propagation. 

\subsection{Local-aligned Score Distilling}

After applying $\enc(x|d;\theta^{\text{(d)}})$ to fine-grained text piece in $d$, we can obtain fine-grained representations, respectively. That is 
\begin{align}
    \vv^{x^j_k} = \enc(x^j_k|d;\theta^{\text{(d)}}), j\in[0, M], k\in[1,K^j],
\end{align}
where $j$ denotes the index of granularity, $M$ denotes the total number of granularity, $i$ denotes the index of text piece in $j$-th granularity, and $K^j$ denotes the number of total text pieces in $j$-th granularity.
Here, $j=0$ denotes the granularity at the document level, leading to $K^0=1$ and $d=x^0_1$. 

Then, we rewrite Eq.(\ref{equ:bi_enc}) to score multi-grained pieces as 
\begin{align}
    \notag s^{(\text{be})}_{j,k} &\coloneqq \gM^{(\text{be})}(q, x^j_k|d;\theta^{(\text{be})}) = <\vu,\vv^{x^j_k}>   \\
    &\coloneqq<\enc(q|\theta^{(\text{q})}), \enc(x^j_k|d;\theta^{(\text{d})})>. \label{equ:bi_enc_multi}
\end{align}
Next, following Eq.(\ref{equ:bi_score_dist}), we can also derive multi-granular score distributions as
\begin{align}
    &\vp^{(\text{be})}_{j,k} \coloneqq P(\rx^j_k|q, \{x^j_{k+}\}\cup\sN^j_k;\theta^{\text{(be)}}) = \label{equ:bi_score_dist_multi} \\
    \notag &~~~\dfrac{\exp(\gM^{(\text{be})}(q, x^j_k|d;\theta^{(\text{be})})/\tau)}{\sum\nolimits_{x'^j_k\in\{x^j_{k+}\}\cup\sN^j_k} \exp(\gM^{(\text{be})}(q, x'^j_k|d;\theta^{(\text{be})})/\tau)},
\end{align}
where $\sN^j_k$ denotes a set of negative samples in $j$-th granularity, which we will dive into in the next sub-section. 

After, we could apply the cross-encoder to each pair of $q$ and $x^j_k$ and its negative pairs for multi-granular distributions. 
It is noteworthy that differing from the bi-encoder, the score between the $q$ and each $x^j_k$ by cross-encoder is based solely on $x^j_k$, independent of the other parts in $d$. 
This is because, in contrast to our bi-encoder that takes global-consistent fine-grained representations to align document-level bottleneck learning, the cross-encoder here aims to provide precise relevance scores to describe $q$-$x^j_k$ relationships exactly. 
Therefore, we can obtain the cross-encoder's relevance scores by
\begin{align}
    s^{(\text{ce})}_{j,k} \coloneqq \gM^{(\text{ce})}(q, x^j_k|\theta^{(\text{ce})}). \label{equ:cross_enc_multi} 
\end{align}
Then, it is also straightforward to get multi-granular score distribution by the cross-encoder, i.e., $\vp^{(\text{ce})}_{j,k} \coloneqq P(\rx^j_k|q, \{x^j_{k+}\}\cup\sN^j_k;\theta^{\text{(ce)}})$. 

Lastly, we can define the training loss of our multi-granular aligned distillation as 
\begin{align}
    L^{\text{(fkd)}} = \sum_{j\in[1,M]} \dfrac{1}{K^j} \sum_{k\in[1,K^j]} \kldiv(\vp^{(\text{be})}_{j,k} \Vert \vp^{(\text{ce})}_{j,k}),
\end{align}
where $\kldiv(\cdot \Vert \cdot)$ denotes the Kullback–Leibler divergence between the two distributions. 
It is remarkable that we do not include $j=0$ here as the document-level relevance is only learned via contrastive learning. 
After replacing $L^{\text{(kd)}}$ in \S\ref{sec:be_learning} with the above $L^{\text{(fkd)}}$, we get the final training loss of our FGD, i.e., 
\begin{align}
    L^{\text{(be)}}_{\theta^{\text{(bi)}}} = \lambda L^{\text{(cl)}} + L^{\text{(fkd)}}. \label{equ:loss_FGD}
\end{align}
Please refer to Figure~\ref{fig:model} for the illustration.

\paragraph{Remark on Overheads. }
The first thought that comes into our mind is that such extensive knowledge distillation from a heavy network will lead to massive training computation overheads. On the side of the student bi-encoder, there is only a little extra computation (i.e., applying the attention pooling multiple times with off-the-shelf attention scores as defined by Eq.(\ref{equ:multi_granular_repre})) in the top layer of the Transformer. 
On the side of the teacher cross-encoder, as the overheads grow quadratically with sequence length (i.e., $\gO(n^2)$), applying cross-encoder to sub-granularity (e.g., passage and sentence) only results in a complexity of $\gO(n\log n)$. Therefore, the complexity brought by calling the cross-encoder is still $\gO(n^2)$. 
Again, we would like to mention that we still use one single bottleneck vector to represent each document instead of multiple vectors \citep{Khattab2021ColBERTv2,Humeau2020polyencoder}, where the multi-granular embeddings serve only as the intermediate for distillations. 

\subsection{Hierarchical Hard Negative Mining}

Hard negative mining has been proven very effective in achieving competitive performance by many previous works \citep{Xiong2021ANCE,Wang2022simlm}. 
Basically, it leverages the best-so-far retriever to retrieve hard examples (i.e., top-relevant documents but not $d_+$) for each query $q$, which are used as negative documents $\sN$ for the next round of retriever training. 

Nonetheless, as formulated in Eq.(\ref{equ:bi_score_dist_multi}), negative text pieces $\sN^j_k$ are needed to sample at each $j$-th granularity. 
Notably, we cannot get the precise gold label(s) at every sub-document granularity $x^j_{k+}$ in Eq.(\ref{equ:bi_score_dist_multi}) except for the gold document $d_+$ (i.e., $j=0$). 
As a weakly-supervised remedy \citep{Yang2022TCL}, we regard each $x^j_k \in d_+$ as a positive text piece during our multi-granular aligned distillation. 
Thereby, we present a simple yet effective hierarchical hard negative mining technique from top to bottom. 
That is, 
\begin{align}
    \sN^j_k = \{x^j_{k-} | x^j_{k-}\sim&\gM^{(\text{be})}(q, x^j_k|d;\theta^{(\text{be})}) \\
    &\wedge x^j_k \in \sN^{j-1}_* \},
\end{align}
where $\sN^{j-1}_*$ denotes all negatives in $(j-1)$-th granularity and $\sN^0 = \sD\setminus\{d_+\}$.

\section{Experiments}

\subsection{Datasets and Evaluation Metrics}
In experiments, we conduct extensive evaluations of our method on two datasets: MS-Marco \citep{Nguyen2016MSMARCO} and TREC Deep Learning 2019 document retrieval (TREC 2019) \citep{Craswell2020TREC19}.
MS-Marco is a widely-used document retrieval dataset, comprising 3.2 million documents, 367 thousand training queries and 5 thousand development queries.
TREC 2019 is a test set in the MS-Marco document ranking task, consisting of 43 queries with more comprehensive labeling.
Following previous works \citep{Ma2022COSTA}, we use official metrics MRR@100 and Recall@100 (R@100) to report the evaluation result on MS-Marco dev. For TREC Deep Learning 2019, we report both nDCG@10 and Recall@100.

\subsection{Pre-training \& Fine-tuning Pipeline}

\begin{figure}[t]
    \centering
    \includegraphics[width=0.85\linewidth]{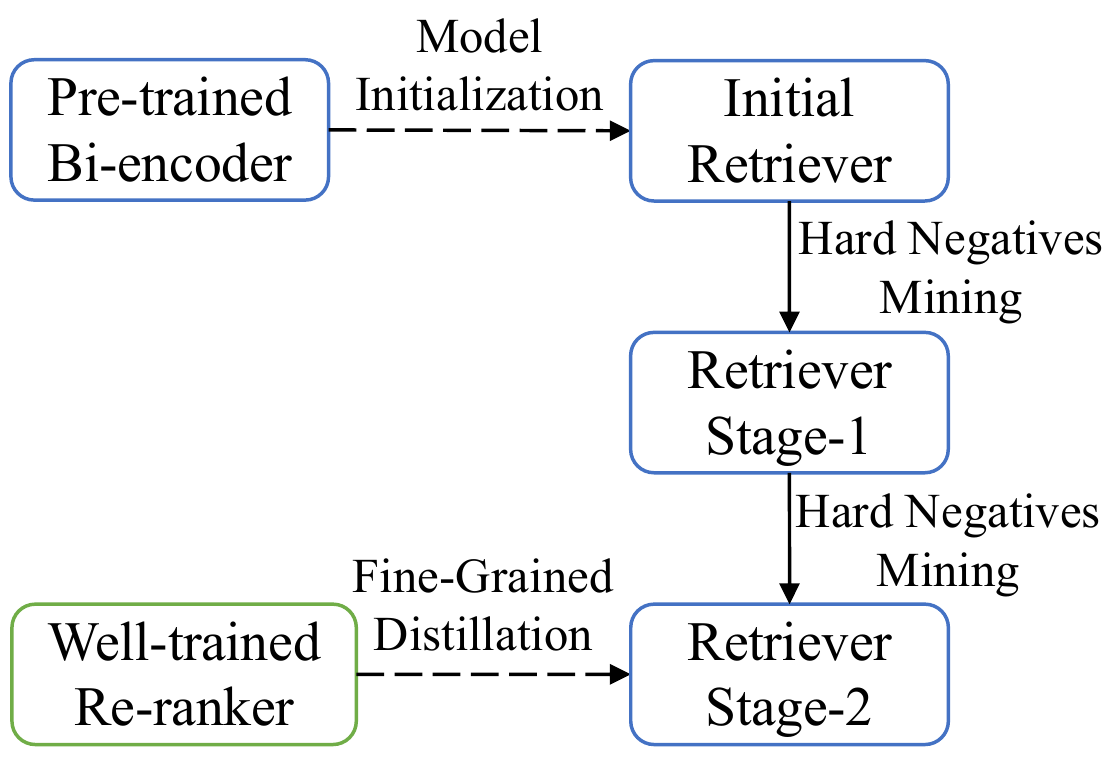}
    \caption{\small The pipeline of our method.}
    \label{fig:pipeline}
\end{figure}

Following previous works \citep{Ma2022COSTA}, we elaborate on our pre-training and fine-tuning pipeline (see a flow chart in Figure~\ref{fig:pipeline}) to achieve the proposed FGD for document retrieval. 

\paragraph{Stage-0: Pre-training.}
Initialing a model by self-supervised pre-training has been proven effective by numerous works \citep{Xiong2021ANCE,Zhan2021STAR-ADORE,Ma2022Bprop,Ma2022COSTA}, which can be categorized into two groups, i.e., \textit{general pre-training} and \textit{corpus-aware pre-training}. 
Specifically, the former is referred to as PLMs that are pre-trained on general corpora by language modeling (e.g., RoBERTa \citep{Liu2019RoBERTa}). Built upon the former, the latter is proposed for continual pre-training on the collection corpus by language modeling and/or pseudo-label training (e.g., coCondenser \citep{Gao2022coCondenser} and SimLM \citep{Wang2022simlm}). 
In this work, we test our framework on both, corresponding to RoBERTa and ED-MLM \citep{Wang2022simlm}. 
In addition, following all previous works in document retrieval \citep{Xiong2021ANCE,Zhan2021STAR-ADORE}, we also conduct a supervised pre-training on passage retrieval by default. 

\paragraph{Stage-1: Warmup Fine-tuning.}

Providing the document-level hard negatives mined by the pre-trained retriever, the first fine-tuning step is based solely on the contrastive learning loss defined in Eq.(\ref{equ:contrastive_learning}) to warm up in retriever  for document retrieval \citep{Zhan2021STAR-ADORE,Wang2022simlm}. 

\paragraph{Stage-2: Continual Fine-tuning.}
Upon the retriever from the warmup stage, the hard negative mining is invoked again for more challenging negatives. In contrast to previous works that merely employ the contrastive learning \citep{Ma2022COSTA}, we apply our proposed FGD here by Eq.(\ref{equ:loss_FGD}) for more competitive results.

\begin{table*}[t] \small
\centering
\begin{tabular}{lcccc}
\toprule
\multicolumn{1}{l}{\multirow{2}{*}{\textbf{Method}}} & \multicolumn{2}{c}{\textbf{MS-MARCO Doc Dev}} & \multicolumn{2}{c}{\textbf{TREC 2019 Doc}} \\ 
\cmidrule(lr){2-3} \cmidrule(lr){4-5}
\multicolumn{1}{c}{}                                 & \textbf{MRR@100}      & \textbf{R@100}     & \textbf{nDCG@10}     & \textbf{R@100}     \\
\hline\midrule
\multicolumn{5}{l}{\it Sparse or lexicon retriever}                                                                                                 \\
\midrule
BM25                                                 & 0.277                 & 0.808              & 0.519                & 0.395              \\
DeepCT \cite{Dai2019DeepCT}                           & 0.320                 & -                  & 0.544                & -                  \\
BestTRECTrad  \cite{Craswell2020TREC19}              & -                     & -                  & 0.549                & -                  \\
\hline\midrule
\multicolumn{5}{l}{\it Dense retriever}                                                                                                  \\
\midrule

ANCE   \cite{Xiong2021ANCE}                     & 0.377                 & 0.894              & 0.610                & 0.273              \\
BERT  \cite{Ma2022COSTA}                                 & 0.389                 & 0.877              & 0.594                & 0.301              \\
STAR   \cite{Zhan2021STAR-ADORE}                       & 0.390                 & 0.913              & 0.605                & 0.313              \\
ICT    \cite{Lee2019ICT}                                & 0.396                 & 0.882              & 0.605                & 0.303              \\
PROP    \cite{Ma2022Bprop}                               & 0.394                 & 0.884              & 0.596                & 0.298              \\
B-PROP  \cite{Ma2022Bprop}                               & 0.395                 & 0.883              & 0.601                & 0.305              \\
SEED    \cite{Lu2021SeedEncoder}                               & 0.396                 & 0.902              & 0.605                & 0.307              \\
RepCONC \cite{Zhan2022RepCONC}                        & 0.399                 & 0.911              & 0.600                & 0.305              \\
JPQ     \cite{Zhan21Jointly}                         & 0.401                 & 0.914              & 0.623                & -                  \\
ADORE+STAR \cite{Zhan2021STAR-ADORE}                   & 0.405                 & 0.919              & 0.628                & 0.317              \\
SeDR  \citep{Chen2022SeDR}                                 & 0.409                 & 0.921              & 0.632                & 0.343              \\
COSTA  \cite{Ma2022COSTA}                                & 0.422                 & 0.919              & 0.626                & 0.320              \\
\midrule
FGD  (ours)                                    & \textbf{0.440}        & \textbf{0.925}     & \textbf{0.635}       & \textbf{0.349}     \\
\bottomrule
\end{tabular}
\caption{Comparison results on MS-Marco and TREC 2019 datasets.}
\label{tab:main}
\end{table*}

\subsection{Implementation Details}

\paragraph{Pre-training Setups.}
We adopt the PLM, RoBERTa-base, as our general pre-trained model. 
Upon this PLM, we conduct a corpus-aware pre-training by following ED-MLM \citep{Wang2022simlm} objective. Specifically, we first make slide windows with a length of 384 and a stride of 64 over the documents from the MS-Marco collection. 
The learning rate is set to $1\times10^{-4}$, the batch size is set to 2048, the number of training epochs is set to 3, and the random seed is set to 42. The other parameters are strictly following \citet{Wang2022simlm}. Such a corpus-aware pre-training procedure takes about 20 hours on eight A100 GPUs.

\paragraph{Fine-tuning Setups.}
The hyperparameters across two-stage document retriever training are shown as follows.
In the first stage, our document retriever initializes from our pre-trained retriever, while the second-stage model initializes from the first-stage retriever.
For model training, We use an Adam optimizer with a learning rate of $3\times10^{-6}$ and a linear warmup strategy with a warmup step of 1,000.
The number of training epochs in the first stage is 2, and that in the second stage is 20.
The weight decay, maximum document length and maximum query length are set to 0.01, 512 and 32, respectively. 
We intercept multiple consecutive 64- and 128-token fragments in documents as corresponding sentences and passages.
The ranker we used for distillation is R2ANKER \citep{Zhou2022r2anker}.
The batch size is set to 64 with 1 positive and 8 negative documents.
The negative documents are sampled with a depth of 100 (i.e., how many top candidates are in the query-relevant negative pool).
In our experiments, the random seed is always set to 42, and we fine-tune document retrievers on eight A100 GPUs. 

\subsection{Main Results}

We compare our method with other strong competitors on MS-Marco and TREC 2019 datasets. The results are shown in Table \ref{tab:main}. From the table, we can see that sparse or lexicon retrievers underperform dense retrievers. The reason is that dense retrievers can find more semantic relevance between documents and queries in contrast to sparse retrieval. Besides, we can observe that our method outperforms other methods and achieves state-of-the-art performance on MS-Marco, which demonstrates the effectiveness of our method. Moreover, it is observed that FGD consistently achieves state-of-the-art performance on TREC 2019, which verifies the effectiveness of FGD again. 

\subsection{Ablation Study}
\begin{table}[t] \small
\centering
\begin{tabular}{lcc}
\toprule
\multirow{2}{*}{\textbf{Method}} & \multicolumn{2}{c}{\textbf{MARCO Dev}} \\ \cmidrule{2-3} 
                                 & \textbf{MRR@100}      & \textbf{R@100}     \\
\midrule
FGD (stg2)  & \textbf{0.440}  & \textbf{0.925}     \\
\midrule
$\Diamond$ FGD w/o pass-distill           & 0.435                 & 0.924              \\
$\Diamond$ FGD w/o sent-distill           & 0.435                 & 0.925              \\
\midrule
$\Diamond$ FGD w/ doc-distill             & 0.436                 & 0.924              \\
$\Diamond$ FGD w/ FG pooling & 0.426 & 0.924 \\ 
\midrule
$\Diamond$ only doc-distill & 0.428 & 0.923 \\
$\Diamond$ w/o ALL   & 0.427   & 0.923              \\
\bottomrule
\end{tabular}
\caption{Ablation study. `FG' denotes `fine-grained', `w/o ALL' is equivalent to `ED-MLM' at stage 2 (stg2). }
\label{tab:abl}
\end{table}
To further investigate the effectiveness of our model, we conduct an ablation study, as shown in Table \ref{tab:abl}.
First, when we respectively remove passage- and sentence-level distillation (i.e., pass- and sent-distill), the performance of our model drops, which verifies their effectiveness. 
In addition, the result of our FGD with document-level distillation exhibits performance loss, showing that document-level distillation and fine-grained distillation are incompatible well.
Moreover, we replace the global-consistent granularity embedding method with mean pooling over token representations (i.e., FG pooling), and the performance drops a lot, which demonstrates the effectiveness of the global-consistent granularity embedding method.

\subsection{Impact of Retriever and Ranker}
\begin{table}[t] \small
\centering
\begin{tabular}{lcc}
\toprule
\multirow{2}{*}{\textbf{Method}}       & \multicolumn{2}{c}{\textbf{MARCO Dev}} \\ \cmidrule{2-3} 
                                       & \textbf{MRR@100}      & \textbf{R@100}     \\
\midrule
FGD (ED-MLM + psg-ranker)                                 & \bf 0.440                 & \bf 0.925              \\
\hline\midrule
\multicolumn{3}{l}{\textit{Replacing the bi-encoder (student) retriever}} \\
\midrule
STAR (stg2)                                 & 0.417                 & 0.914              \\
FGD (STAR as student)                      & 0.430                 & 0.915   \\
\hline\midrule
\multicolumn{3}{l}{\textit{Replacing the cross-encoder (teacher) reranker}} \\
\midrule
ED-MLM (stg2)                              & 0.427                 & 0.923              \\
FGD (doc-ranker as teacher) & 0.438                 & 0.923              \\
\bottomrule
\end{tabular}
\caption{Results with different bi-encoder and cross-encoder.}
\label{tab:diff}
\end{table}
We replace different teacher cross-encoder and student bi-encoder in our method to evaluate their impact.
From the table, our method achieves significant improvements (i.e., 0.417 to 0.430 on MRR\@10) with STAR \cite{Zhan2021STAR-ADORE} as the student bi-encoder.
Moreover, we can see that FGD (STAR as student) has a performance drop compared to FGD (ED-MLM + psg-ranker). 
The reason is that STAR is trained from a RoBERTa \citep{Liu2019RoBERTa} (i.e., a general pre-trained encoder), while ED-MLM is a corpus-aware pre-trained encoder in MS-Marco. 
In addition, our method with a document ranker \citep{Gao2021ranker}, i.e., FGD (doc-ranker as teacher),  also achieves significant improvements, which demonstrates the effectiveness of our method again.
\begin{figure*}[t]
\centering
\includegraphics[width=0.31\linewidth]{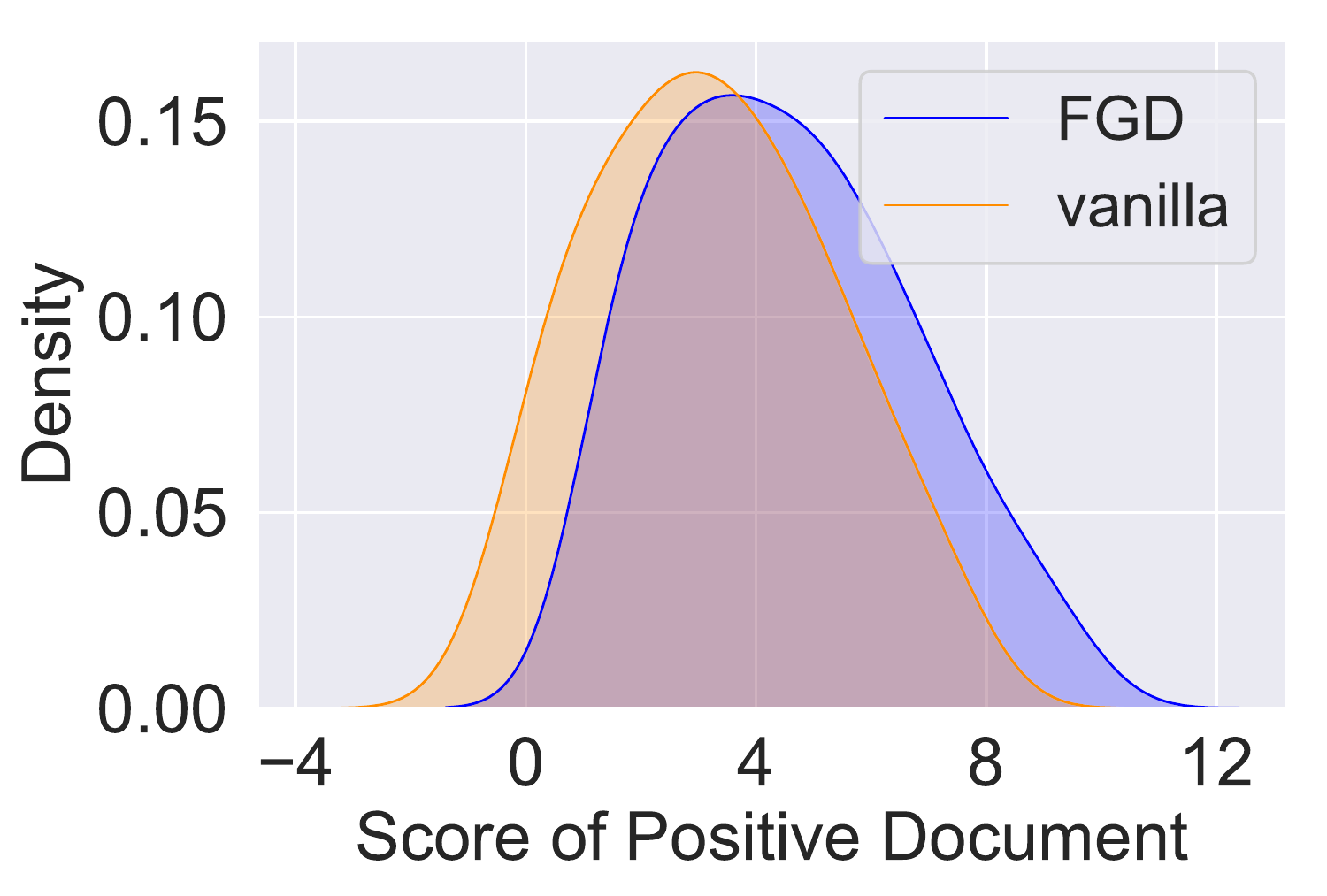}
\includegraphics[width=0.31\linewidth]{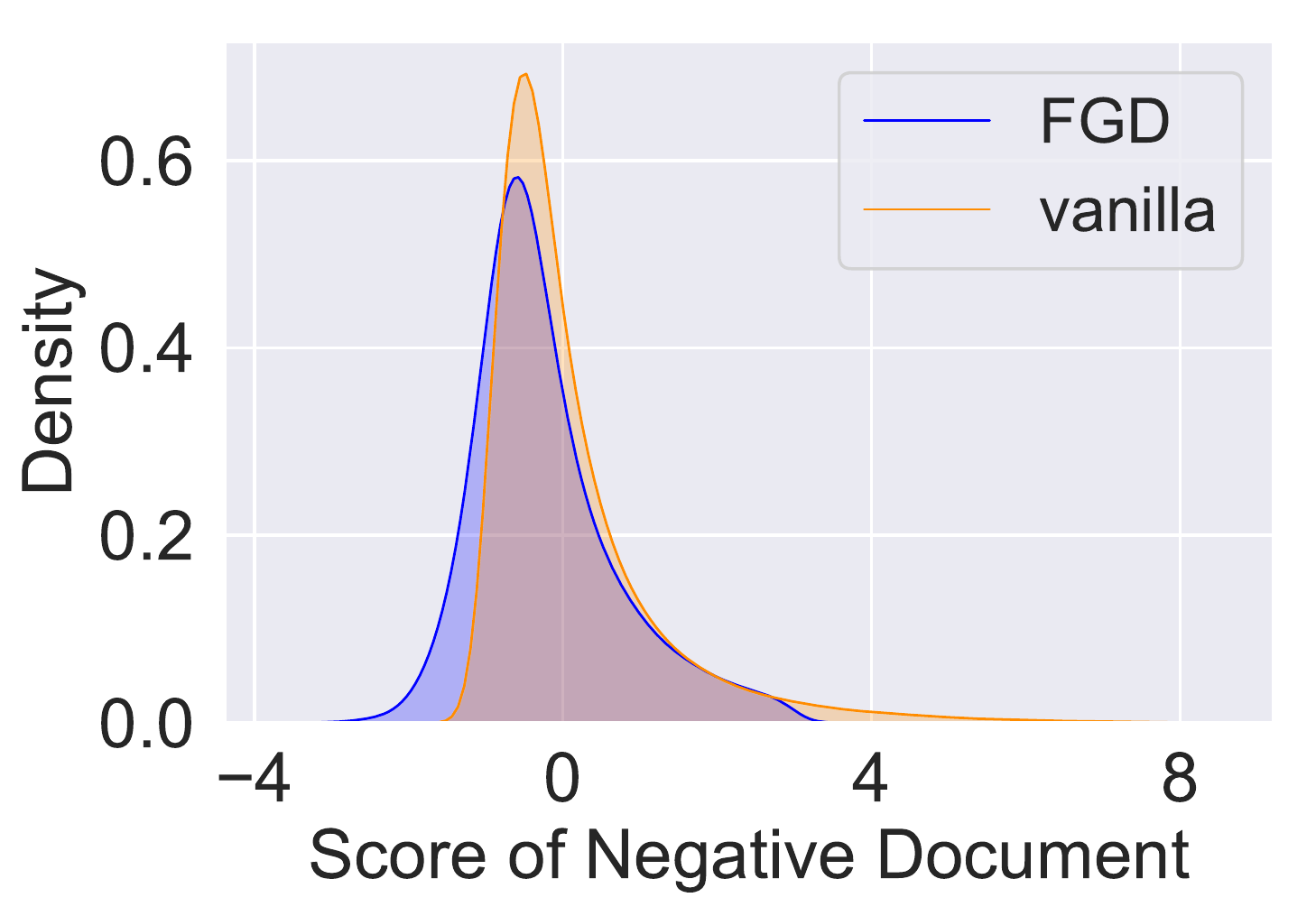}
\includegraphics[width=0.31\linewidth]{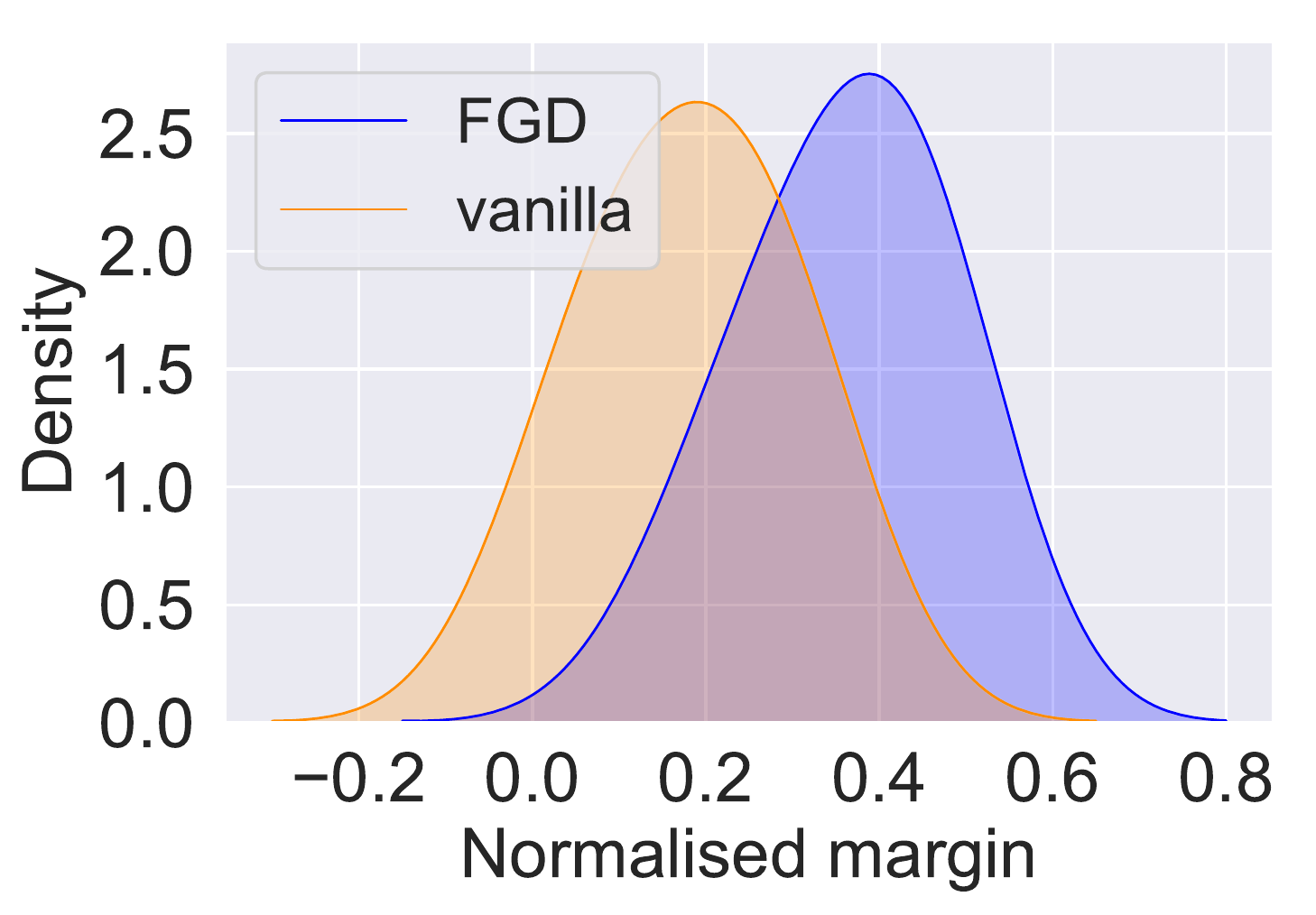}
\caption{\small Comparison of our FGD (left) and vanilla document-level distillation (middle) model predictions on the MS-Marco test set; and the margins (right) of the FGD and vanilla.}
\label{fig:distribution}
\end{figure*}

\subsection{Impact of Fine-Grained Distillation}
In the Figure \ref{fig:distribution}, we show the prediction distributions of our method and vanilla document-level distillation model on the MS-Marco test set. 
From the left and middle of the figure, we can observe that our method makes a greater distinction between positive and negative samples, which shows that our method has a stronger ability to distinguish positive and negative. 
In right of the figure, we calculate normalised margins between the positive and negative pairs based on $(s(q, d^+) - s(q, d^-)) / \beta$, where $\beta$ is the maximal score range.
We can see that our method has greater normalised margins, which further demonstrates that our method is more discriminative between positive and negative documents.

\subsection{Multi-Vector Retrieval}
\begin{figure}[t]
\centering
\includegraphics[width=0.73\linewidth]{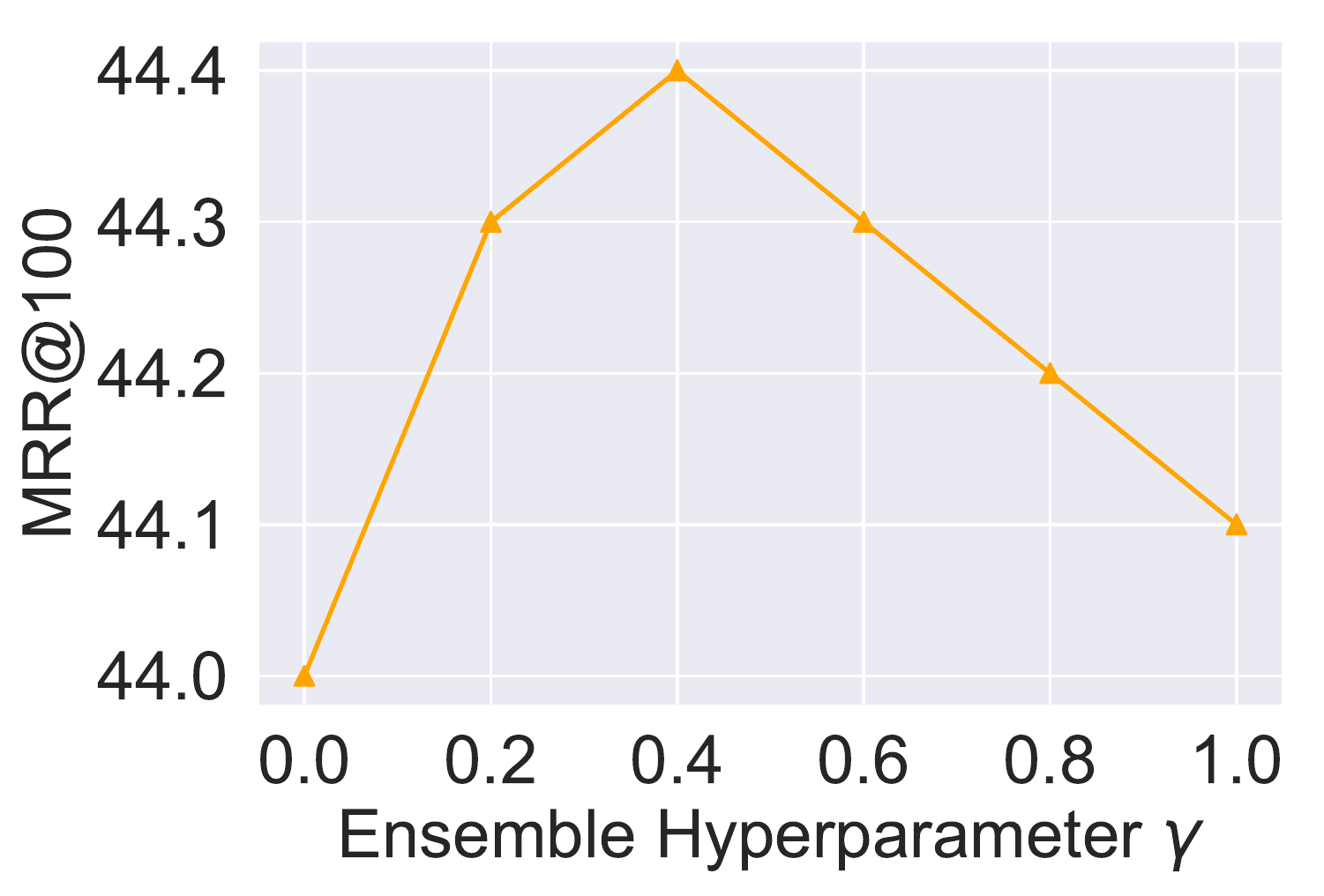}
\caption{\small Change of ensemble hyperparameter $\gamma$ for multi-vector retrieval.}
\label{fig:rate}
\end{figure}

\begin{table}[t]\small
\centering
\begin{tabular}{lcc}
\toprule
\multirow{2}{*}{\textbf{Method}} & \multicolumn{2}{c}{\textbf{MARCO Dev Doc}} \\\cmidrule{2-3}
                                          & \textbf{MRR@100}      & \textbf{R@100}     \\\midrule
Previois SoTA  & 0.422        & 0.919     \\
FGD    & 0.440        & 0.925     \\
FGD + multi           & \textbf{0.444}        & \textbf{0.926}     \\
\bottomrule
\end{tabular}
\caption{\small Boosting FGD with multi-vector (i.e., `multi') retrieval. }
\label{tab:ensem}
\end{table}
Since our method can derive multi-granularity (i.e., document, passage, sentence) representations, we leverage an ensemble hyperparameter $\gamma$ to integrate relevances of query-document, query-passage and query-sentence to investigate the impact on document retrieval, i.e., 
\begin{align}
s^{(be)} := s^{(be)} + \gamma \times (\max(s^{(be)}_{1,i}) + \max(s^{(be)}_{2,j}))
\end{align}
where $s^{(be)}$ is the relevance of query-document. $\max(s^{(be)}_{1,i})$ and $\max(s^{(be)}_{2,j})$ denote maximum relevances of query-passage and query-sentence, and passages and sentences are in the document. From the Figure \ref{fig:rate}, we can observe that when the ensemble hyperparameter $\gamma$ is 0.4, the performance reaches the peak and achieves a new state-of-the-art, as shown in Table \ref{tab:ensem}.

\subsection{Fine-Grained Representation Derivation}
\begin{table}[t]\small
\centering
\begin{tabular}{lcc}
\toprule
\multicolumn{1}{c}{\multirow{2}{*}{\textbf{Method}}} & \multicolumn{2}{c}{\textbf{MARCO Dev Doc}} \\\cmidrule{2-3}
\multicolumn{1}{c}{}                                 & \textbf{MRR@100}      & \textbf{R@100}     \\\midrule
FGD (global-consistent)                                        & 0.440                 & \bf 0.925              \\
- FGD w/ RGAT                                  & \bf 0.441                 & \bf 0.925              \\ 
- FGD w/ FG pooling               & 0.426                 & 0.924              \\
\bottomrule
\end{tabular}
\caption{\small Comparisons on MS-Marco dev w.r.t. different methods to derive fine-grained representations.}
\label{tab:rgat}
\end{table}
Apart from the global-consistent granularity embedding method, we propose two fine-grained representation derivation methods: FG pooling and RGAT.  
FG pooling refers to using a mean pooling operation to aggregate the token representations corresponding to sentences/passages as a sentence/passage representation.
RGAT means constructing a graph by taking representations of sentences, passages and documents from FG pooling as nodes and their relation as edges, then updating their representations using a relational graph attention network. 
As shown in Table \ref{tab:rgat}, the results show that FGD with FG pooling has no gain.
In contrast, FGD with RGAT can achieve similar performance to FGD with global-consistent granularity embedding but needs to introduce more model parameters.

\section{Related Work}

\paragraph{Retriever Training with Distillation.}
In large-scale retrieval, there are two paradigms of mainstream ad-hoc retrieval (a.k.a first-stage retrieval), i.e., dense retrieval \cite{Gao2021Condenser} and sparse retrieval \cite{Shen22LexMAE}.
In contrast to sparse retrieval exploiting almost unlearnable BM25 or language models for term-based retrieval, dense retrieval aims to encode query-entry pairs into dense vectors, potentially finding more semantic relevance between entries and queries.
Therefore, a lot of dense retrieval methods are proposed based on the prevalent PLM-based bi-encoder structure \cite{Gao2021Condenser,Lu22ERNIE}.
To improve dense passage retrieval, a recent trend is to conduct distillation from a cross-encoder-based ranker to a dense retriever, where the ranker can be well-trained in advance \citep{Lin2021TCT,Zhou2022r2anker} or updated along with the bi-encoder \citep{Zhang2021AR2}.
In contrast to the conventional setting, distillation in retrieval does not focus on model compression but aims to distill features from different retriever architectures to learn knowledge from different semantic perspectives \cite{menon2021indefense}.
In distillation in retrieval, a well-trained ranker is widely used as the teacher model to produce weak labels on large-scale unlabeled query-document pairs \cite{Ren2021RocketQAv2,Zhang2021AR2,Lu22ERNIE}.
To investigate the effectiveness of distillation, \citet{menon2021indefense} conduct a study on the gap between cross-encoders and bi-encoders and deduce an empirical conclusion that bi-encoders are overfitting to the training set. 
In the study, distillation from the cross-encoder to the bi-encoder has been proven effectively alleviate overfitting on bi-encoders.
However, these methods only investigate how to improve passage retrievers by distillation.

\paragraph{Multi-granular Representation Learning.}
Since encoding of long documents suffers from the \textit{scope hypothesis} that a long document may cover multiple topics \citep{Robertson09Probabilistic}, there is an information bottleneck of long documents' representation.
To break the information bottleneck, many efforts have been made to produce multi-granular representations \citep{Liu2022Multi-Granularityxmodal,Zheng2020DocGNN}.
Previous methods directly apply mean-pooling over contextual embeddings for different granularity \citep{Liu2022Multi-Granularityxmodal}. However, our pilot experiments prove it is less effective.
To obtain multi-granular representations for documents, \citet{Zheng2020DocGNN} construct a graph among different semantic units (e.g., document, passage, sentence) and leverage relational graph attention networks to derive representations of different semantic units over the graph.
However, these methods fail to derive global-consistent representations across granularity. 

\paragraph{Multi-granular Distillation.}
Recently, prevailing methods of distillation \cite{menon2021indefense,Ren2021RocketQAv2} focus on transferring text knowledge from mono-granularity language units (e.g., passage, sentence).
However, mono-granular knowledge usually fails to represent the whole semantics of a text, i.e., losing some vital knowledge.
To address this problem, \citet{Liu2022Multi-Granularity} propose multi-granularity knowledge distillation to exploit information of multi-granularity language units for model compression.
Despite their success, these methods rely on the same structure between teacher and student models, which fails to transfer knowledge across architectures.
Motivated by \citet{Liu2022Multi-Granularity}, we propose a local-aligned score distilling across architectures for document retrieval.

\paragraph{Hard Negative Mining.}
Mining negative samples for retriever training have been proven very effective in achieving competitive performance by many previous works \citep{Xiong2021ANCE,Zhan2021STAR-ADORE,Gao2022coCondenser,Wang2022simlm,Shen22LexMAE}. 
For example, \citet{Huang2020Embedding} randomly sample documents from a document collection as negatives to train a retriever.
To efficiency of retriever training, \citet{Zhan2020RepBERT} leverage in-batch negative training that a query regards other queries' negatives in the same mini-batch as negatives.
In addition, increasing the number of random negatives in the mini-batch is proven effective \citet{Qu2021RocketQA}.
To mine more effective negatives, many works adopt hard negative sampling for retriever training \citep{Xiong2021ANCE,Qu2021RocketQA,Ren2021RocketQAv2}.
BM25 top documents as hard negatives are widely used in many works \citep{Gao2020Complementing,Karpukhin2020DPR}.
To mine harder negatives, \citet{Guu2020RetriAugLM} apply BM25 negatives to train a warm-up retriever, and then use the retriever to retrieve the top documents as hard negatives during training. 
However, these sampling methods are needed to sample at different granularity for multi-granular distillation.
Thereby, we present a simple yet effective hierarchical hard negative mining technique from top to bottom. 

\section{Conclusion}

In this work, we propose a new knowledge distillation framework for long-document retrieval, which is called fine-grained distillation (FGD). 
Integrated with the hierarchical hard negative mining technique, the proposed framework is able to produce fine-grained representations consistent with the global document-level one and then distill multi-granular score distributions from a heterogeneous cross-encoder. 
The proposed learning framework will not affect the long-document retrieval procedure in terms of both retrieval paradigm and efficiency, but only increases limited computation overheads during training. 
The experiment results show that the proposed framework can achieve a state-of-the-art quality in document retrieval and is compatible enough with a broad spectrum of baseline choices in terms of both the bi-encoder student and the cross-encoder teacher.

\bibliography{ref}
\bibliographystyle{acl_natbib}

\end{document}